\documentclass[12pt]{article}
\usepackage{epsfig}
\usepackage{amssymb}
\usepackage{amsmath}
\usepackage{amsfonts}
\usepackage{graphicx}
\usepackage{mathrsfs}
\usepackage[dvips]{color}
\usepackage{multirow}

\newcommand{\bsigma}{\boldsymbol{\sigma}}

\newcommand{\R}{\mathbb{R}}
\newcommand{\C}{\mathbb{C}}
\newcommand{\Z}{\mathbb{Z}}

\newcommand{\fa}{\mathfrak{a}}
\newcommand{\fb}{\mathfrak{b}}

\newcommand{\fz}{\mathfrak{z}}

\newcommand{\fK}{\mathfrak{K}}

\newcommand{\bg}{\mathbf{g}}

\newcommand{\bk}{\mathbf{k}}

\newcommand{\bm}{{\mathbf{m}}}

\newcommand{\bx}{\mathbf{x}}

\newcommand{\bzero}{\mathbf{0}}

\newcommand{\bH}{\mathbf{H}}
\newcommand{\bI}{\mathbf{I}}

\newcommand{\bM}{\mathbf{M}}

\newcommand{\bU}{\mathbf{U}}

\newcommand{\cK}{\mathcal{K}}

\newcommand{\be}{\begin{equation}}
\newcommand{\ee}{\end{equation}}
\newcommand{\bea}{\begin{eqnarray}}
\newcommand{\eea}{\end{eqnarray}}
\newcommand{\nn}{\nonumber}

\newcommand{\ed}{\end{document}}

\newcommand{\bi}{\begin{itemize}}
\newcommand{\ei}{\end{itemize}}

\newcommand{\bce}{\begin{center}}
\newcommand{\ece}{\end{center}}

\newcommand{\sT}{\mathscr{T}}

\newcommand{\bcK}{{\boldsymbol{\cK}}}

\newcommand{\tE}{{\mbox{\tiny$\rm E$}}}
\newcommand{\tst}{{\mbox{\tiny$\rm S.T.$}}}
\newcommand{\hx}{{\widehat x_0}}

\begin{document}

\title{Transmission of low-energy scalar waves through a traversable wormhole}

\author{Bahareh Azad\thanks{E-mail address: b.azad@alumni.iut.ac.ir}~,
Farhang Loran\thanks{E-mail address: loran@iut.ac.ir}~, and
Ali~Mostafazadeh\thanks{E-mail address:
amostafazadeh@ku.edu.tr}\\[6pt]
$^{*,\dagger}$Department of Physics, Isfahan University of Technology, \\ Isfahan 84156-83111, Iran\\[6pt]
$^\ddagger$Departments of Mathematics and Physics, Ko\c{c}
University,\\  34450 Sar{\i}yer, Istanbul, Turkey}

\date{ }
\maketitle

\begin{abstract}

We study the scattering of low-energy massless and massive minimally coupled scalar fields by an asymptotically flat traversable wormhole. We provide a comprehensive treatment of this problem offering analytic expressions for the transmission and reflection amplitudes of the corresponding effective potential and the absorption cross section of the wormhole. Our results, which are based on a recently developed dynamical formulation of time-independent scattering theory, apply to a large class of wormhole spacetimes including a wormhole with a sharp transition, the Ellis wormhole, and a family of its generalizations.




\end{abstract}

\section{Introduction}

Wormholes have been a focus of attention for several decades. Since the pioneering work of Einstein and Rosen \cite{ER-bridge}, their study was mainly directed at addressing two basic questions: How can
a traversable wormhole be constructed, and how can such a wormhole be probed? Refs.~\cite{censor1,censor2,censor3,censor4,censor5,Hochberg:1997wp,interstellar} study the existence of stable  traversable wormholes in general relativity. Under the assumption that our universe is a four dimensional asymptotically flat spacetime, they provide arguments against their natural existence. These do not however exclude the possibility of their occurrence in other theories of gravity
\cite{Bhawal:1992sz,Kanti:2011jz,Restuccia:2020wls,Korolev:2020ohi,higher1,higher2,higher3,higher4,Blazquez-Salcedo-2020czn}. Nor do they exhaust the possibility of the presence of wormholes with a sophisticated structure as proposed in Refs.~\cite{creation1,creation2,creation3}.

A physically more relevant task is to seek for the consequences of the existence of wormholes and to devise appropriate means for their detection. Refs.~\cite{Lensing1,Lensing2} suggest using the gravitational lensing effects for probing wormholes, while Refs.~\cite{Dai19,Krasnikov19,Dai-2019nph,Simonetti-2020vhw} study the effects of the gravitational pull of the objects residing in the vicinity of one of the mouths of a wormhole on the trajectory of objects hovering around the other mouth. These studies together with the spectroscopic techniques \cite{Spectroscopy} should, in principle, offer useful methods to decide whether a given celestial body is a traversable wormholes \cite{Cardoso:2016rao}.

The study of the effects of a background wormhole geometry on the propagation of waves offer alternative means for probing wormholes \cite{clement-1984,kar-1994,kar-1995,propagate1,propagate4}.
Pursuing this line of research, the authors of Ref.~\cite{Doroshkevich2008} have conducted a numerical study of the process of the transmission of a pulse through a wormhole and its back-reaction to the wormhole geometry. Ref.~\cite{Konoplya10}, uses the WKB formula for finding the quasinormal modes of a wormhole spacetime and the corresponding S-matrix. Ref.~\cite{Konoplya18} shows that the WKB approximation can be used to compute the shape function of a spherically symmetric traversable wormhole near its throat from its high-frequency quasinormal modes.

The recent construction of a class of wormholes in four-dimensional asymptotically flat spacetimes, which are sufficiently stable to be considered traversable \cite{Maldacena18,Marolf19}, has provided added incentive for the study of wormholes. These wormholes are nevertheless very fragile and can only be probed using very low-energy waves. This provides our basic motivation for the present investigation.

The scattering problem for scalar and electromagnetic waves propagating in a background wormhole spacetime has been perviously considered for specific cases in Refs.~\cite{clement-1984,kar-1994,kar-1995}. Ref.~\cite{clement-1984} considers a scattering set-up where both the source of the incident waves and the detector reside in one of the two universes connected by the wormhole while the Dirichlet boundary condition imposed at the wormhole's throat prohibits the transmission of the wave to the other universe. The authors of Refs.~\cite{kar-1994,kar-1995} do not impose such a boundary condition, but for technical reasons their results apply to a specific class of wormholes and a discrete set of energies of the incident wave that makes a detailed study of the transmission of low-energy waves through the wormhole intractable. 

Another problem of direct relevance to the study of the scattering of waves by a wormhole is that of the determination of its absorption cross section. Ref.~\cite{Lima-2020} addresses this problem for massless scalar waves propagating in a family of spacetimes that interpolate between a Schwarzschild blackhole and a particular wormhole. The numerical results reported in  Refs.~\cite{Lima-2020} show that, similarly to the case of a Schwarzschild blackhole \cite{Unruh-1976}, as the energy of the incident wave approaches zero, the absorption cross section of the wormhole tends to a positive value. The behavior of the absorption cross section of more general blackholes for low-energy massless scalar fields has been investigated in Refs.~\cite{das-1997,Higuchi-2001,Magalhes-2020}.

In contrast to earlier investigations of the scattering problem for waves in a wormhole background, in the present article we offer a comprehensive analytic treatment of the low-energy scattering of massless as well as massive scalar waves propagating in a large class of wormhole spacetimes. Our basic tool is a recently developed dynamical formulation of time-independent scattering theory in one dimension \cite{ap-2014,pra-2014a} which provides a practical prescription for computing the reflection and transmission amplitudes to the leading and next-to-leading order terms in powers of the wavenumber \cite{p162}.

\section{Scalar fields in a wormhole spacetime}

Consider a pair of Minkowski spacetimes $M^\pm$, and choose a common time variable $t$ and spherical coordinates $(r_\pm,\vartheta,\varphi)$ to parameterize the spatial slices $\Sigma^\pm_t$ of $M^\pm$ corresponding to $t={\rm constant}$. Let $x:=\pm r_\pm$. Then $(x,\vartheta,\varphi)$ with $x\in\R$ mark the points of the disjoint union of $\Sigma^-_t$ and $\Sigma^+_t$ with their origins identified. This shows that the Minkowski line element,
	\be
	ds^2=-dt^2+dx^2+x^2d\Omega^2,
	\label{e1}
	\ee
with $d\Omega^2:=d\vartheta^2+\sin^2\vartheta d\varphi^2$, and the radial coordinate $x$ taking both negative and positive values describes a spacetime $M$ consisting of a pair of Minkowski spacetimes, namely $M^\pm$, that are glued together along their $t$-axes ($r_\pm=0$.)

\begin{figure}
  	\begin{center}
 	\includegraphics[scale=0.2]{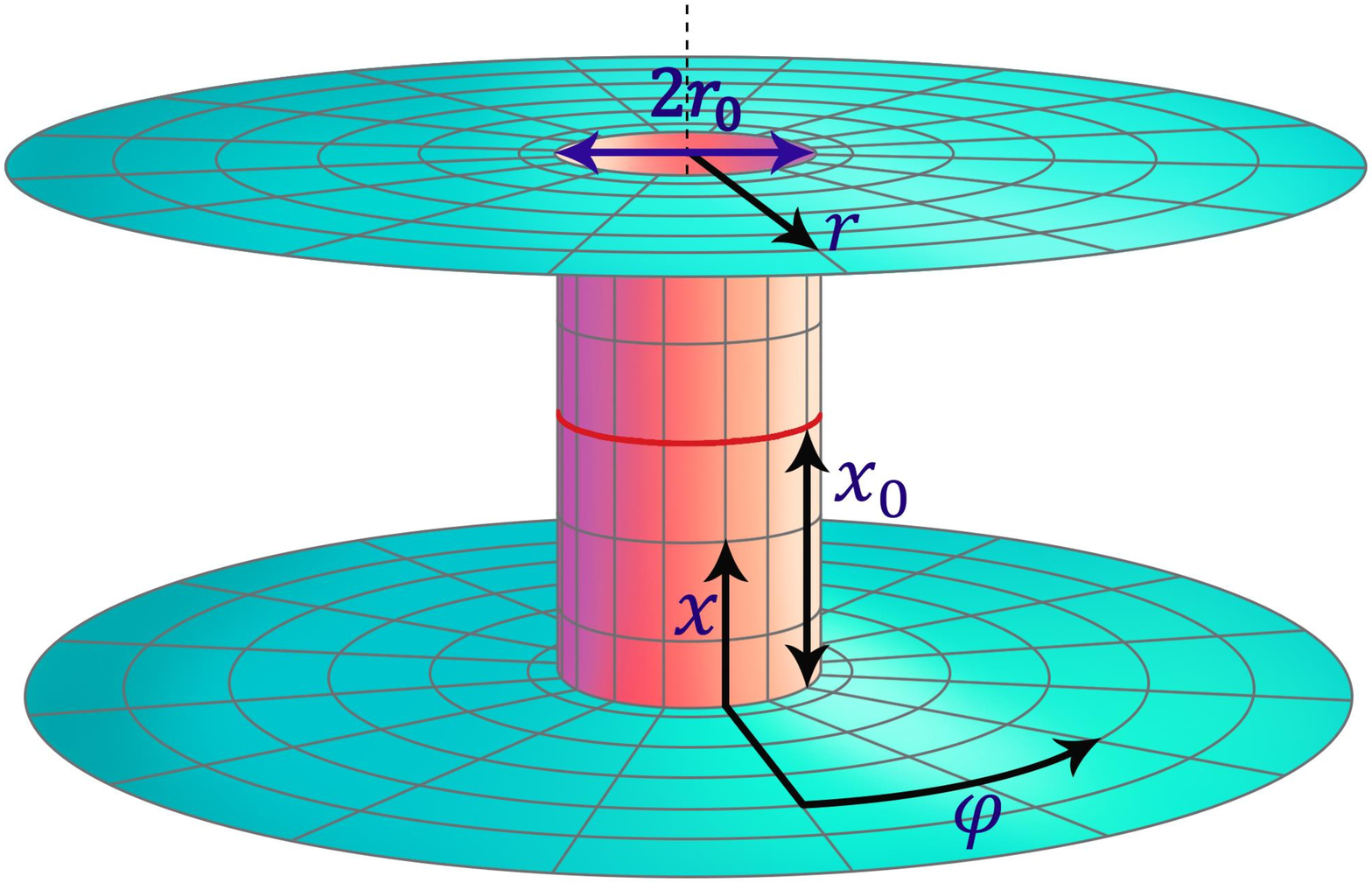}\hspace{1cm}
 	\includegraphics[scale=0.27]{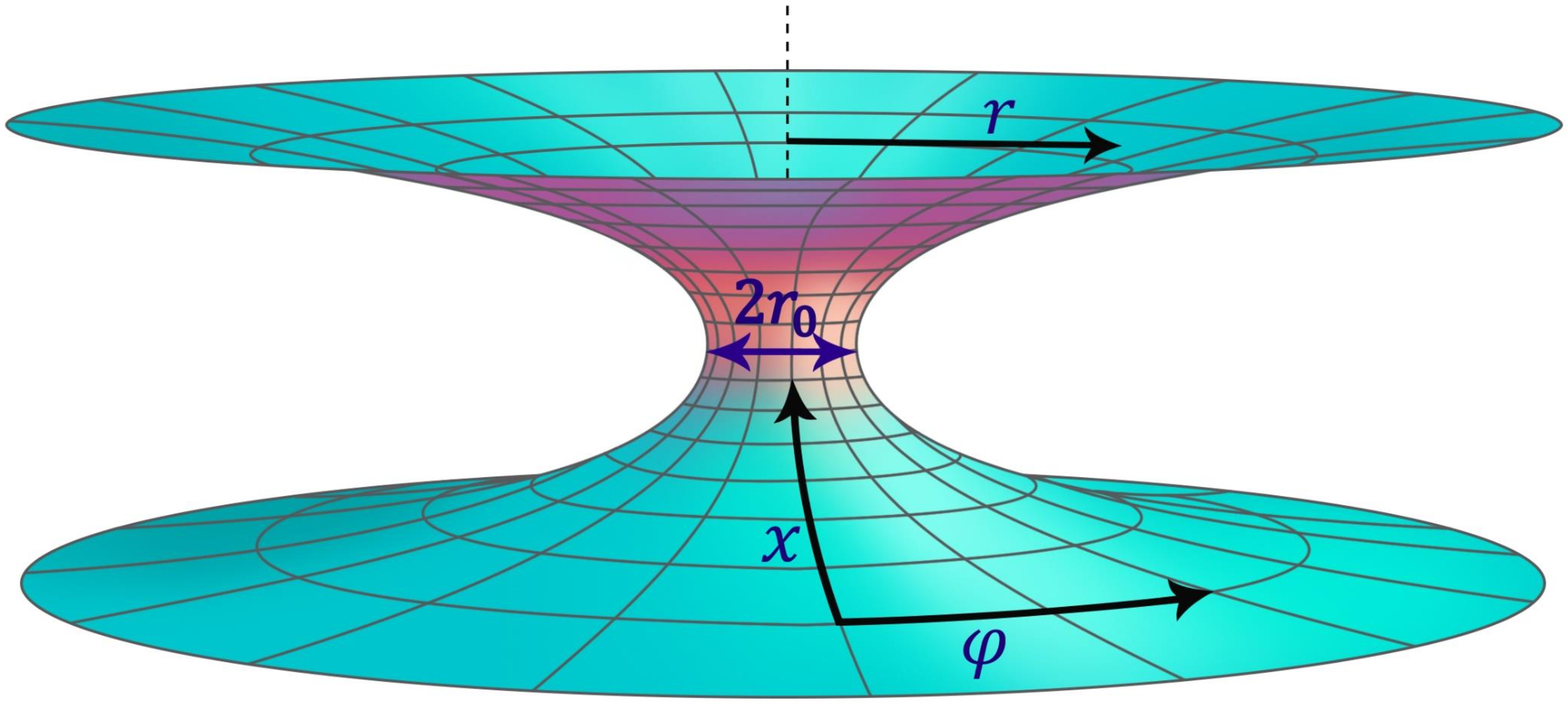}
  	\end{center}
  	\vspace{-12pt}
  	\caption{Schematic demonstration of a wormhole with sharp transition (on the left) and the Ellis wormhole (on the right).}
	\label{fig1}
	\end{figure}

Roughly speaking $M$ corresponds to a static wormhole with a sharp transition \cite{interstellar}, if we replace the $t$-axes along which $M^\pm$ are glued in the above construction with the cylinders $C^\pm=\cup_{t\in\R}(\{t\}\times B_t^\pm)$ having spatial slices $B_t^\pm:=\big\{(r_\pm,\vartheta,\varphi)\in\Sigma^\pm_t\:|\:r_\pm\leq r_0\:\big\}$, for some $r_0\in\R^+$. 
The exterior of the wormhole is the disjoint union of the flat spacetimes obtained by removing $C^\pm$ from $M^\pm$. Its points are parameterized by the coordinates $(t,x,\vartheta,\varphi)$ with $|x|>r_0$. The interior of the wormhole is parameterized by the same coordinates with $|x|<r_0$. The interior and exterior of the wormhole are then glued along their boundaries $x=\pm r_0$. See Fig.~\ref{fig1}. The more realistic variant of this construction, which would correspond to its $r_0\to\infty$ limit, is the one in which $M^\pm$ are asymptotically flat spacetimes and the role of $B_t^\pm$ is played by their suitable deformations such that $M$ is a smooth manifold. A typical example is the Ellis wormhole \cite{Ellis-73} which is also depicted in Fig.~\ref{fig1}.

In general a static wormhole can be more precisely described by a line element of the form,
	\be
	ds^2=-p(r)^2 dt^2+q(r)^2 dr^2+r^2d\Omega^2,
	\label{WH1}
	\ee
where $r$ is a radial spherical coordinate, and $p$ and $q$ are real-valued functions \cite{Morris-1988}. We demand that the latter take strictly positive values, so that the wormhole does not have an event horizon. Let us perform a coordinate transformation, $r\to x$, where $x$ takes values in the whole real line and satisfies
	\be
	dx=q(r)dr.
	\label{ODE}
	\ee
Then (\ref{WH1}) reads
	\be
	ds^2=-p(r)^2 dt^2+dx^2+r^2d\Omega^2.
	\label{WH}
	\ee
In terms of $x$, the exterior of the wormhole correspond to $|x|>x_0$, where $x_0$ fulfills $r(x_0)=r_0$. We can view the latter as an initial condition for the differential equation (\ref{ODE}). Therefore, $(x_0,r_0)$ are the numerical data determining the wormhole for given functions $p$ and $q$. For example, setting
	\begin{align}
	&p(r)=1, && r(x)=\sqrt{x^2+r_0^2},
	\label{ellis-wh}
	\end{align}
in (\ref{WH}), we arrive at the line element for the Ellis wormhole \cite{Ellis-73}, while
	\begin{align}
	&p(r)=1, && r(x)=\left\{\begin{array}{ccc}
	r_0 &{\rm for} & |x|\leq x_0,\\
	|x|-x_0+r_0 &{\rm for} & |x|> x_0,\end{array}\right.
	\label{sharp}
	\end{align}
correspond to a wormhole with a sharp transition \cite{interstellar}. Note that the choice $x_0=r_0=0$ in (\ref{sharp}) reduces (\ref{WH}) to the line element of the degenerate wormhole (\ref{e1}).

Now, consider a minimally coupled complex scalar field $\Phi$ of mass $m$. By definition, it satisfies
    \be
    (-g)^{-\frac{1}{2}} \partial_\mu\left[(-g)^{\frac{1}{2}} g^{\mu\nu}\partial_\nu \Phi\right]-m^2\Phi=0,
    \label{scalar-field}
    \ee
where $g_{\mu\nu}$ are local components of the metric tensor associated with the line element (\ref{WH}), $g^{\mu\nu}$ are the entries of the inverse of the matrix $\bg:=[g_{\mu\nu}]$, and $g:=\det\bg$.

Because the wormhole spacetime is asymptotically flat, for $|x|\to\infty$ the solutions of the field equation (\ref{scalar-field}) tend to those of a free Klein-Gordon equation propagating in a Minkowski spacetime. This shows that (\ref{scalar-field}) admits asymptotically plane-wave solutions that we can employ to define a proper scattering problem. As a first step towards addressing this problem, we use the $t$-independence and spherical symmetry of the metric to construct time-harmonic solutions of (\ref{scalar-field}) that have the following form.
	\be
    	\Phi(t,x,\vartheta,\varphi)=e^{-i\omega t} e^{u(x)}{Y_l^m}(\vartheta,\varphi)\psi(x)
	\label{basic-sols}
    	\ee
where $\omega$ is a positive real parameter,
	\be
	u(x):=-\frac{1}{2}\ln\left[r^2p(r)\right]\Big|_{r=r(x)},
	\label{u-def}
	\ee
$Y_l^m$ are the spherical harmonics, ${l}\in\{0,1,2,\cdots\}$, $m\in\{0,\pm 1,\pm 2,\cdots,\pm l\}$, and $\psi$ is a complex-valued function. Substituting (\ref{basic-sols}) in (\ref{scalar-field}), we can identify the latter with a solution of the time-independent Schr\"odinger equation,
    \be
    -\psi''(x)+v(x)\psi(x)=k^2\psi(x),
    \label{sch-eq}
    \ee
where a prime stands for derivation with respect to $x$, $k:=\sqrt{\omega^2-m^2}$, and
    \be
    v(x):=\left.\left\{\frac{{l}({l}+1)}{r^2}+{u'(x)}^2-u''(x)+(k^2+m^2)\left[1-\frac{1}{p(r)^{2}}\right]\right\}\right|_{r=r(x)}.
    \label{potential}
    \ee
This argument reduces the scattering problem defined by the field equation (\ref{scalar-field}) to a problem of potential scattering in non-relativisitic quantum mechanics.

\begin{figure}
  	\begin{center}
 	\includegraphics[scale=0.5]{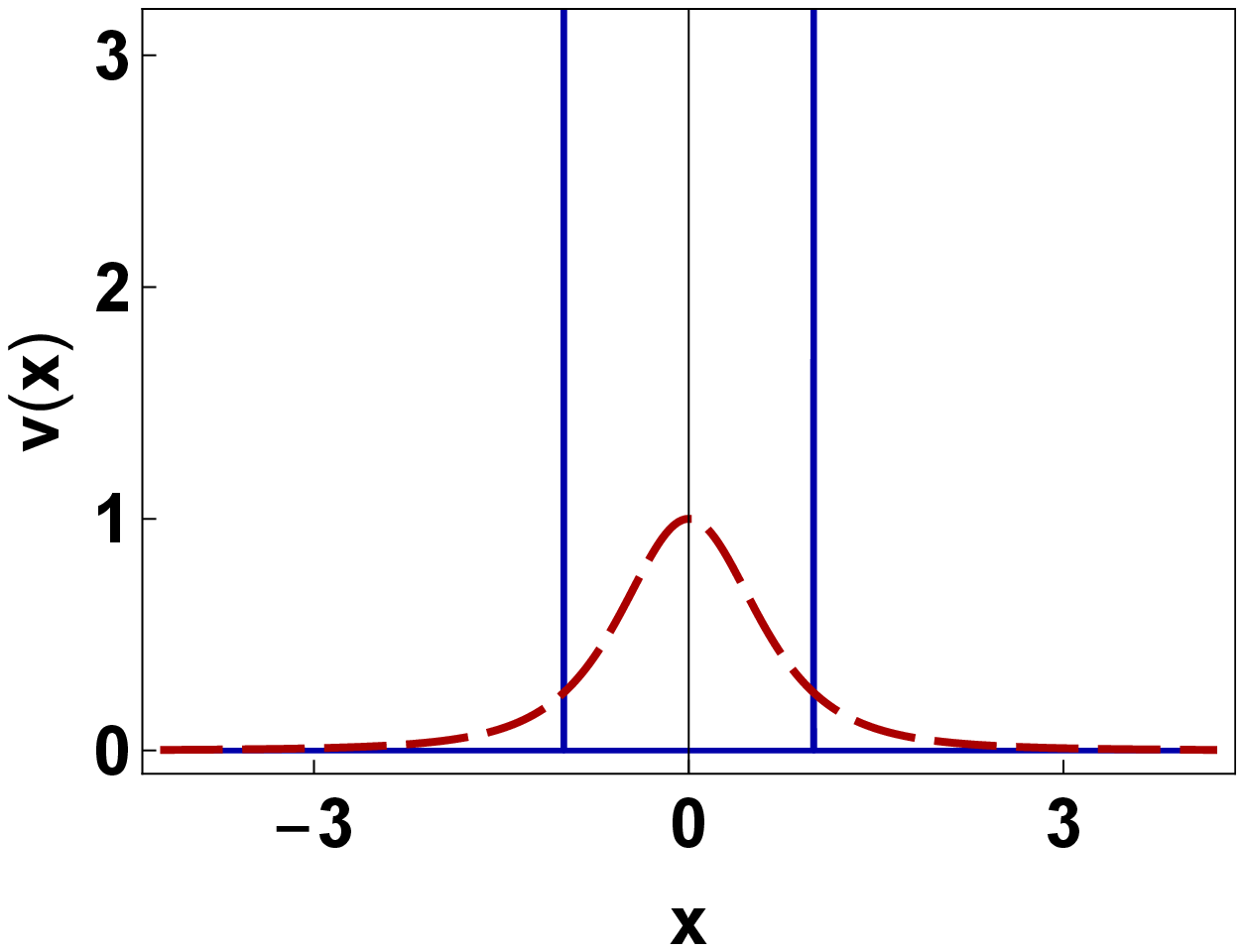}
	\hspace{1cm}
	\includegraphics[scale=0.5]{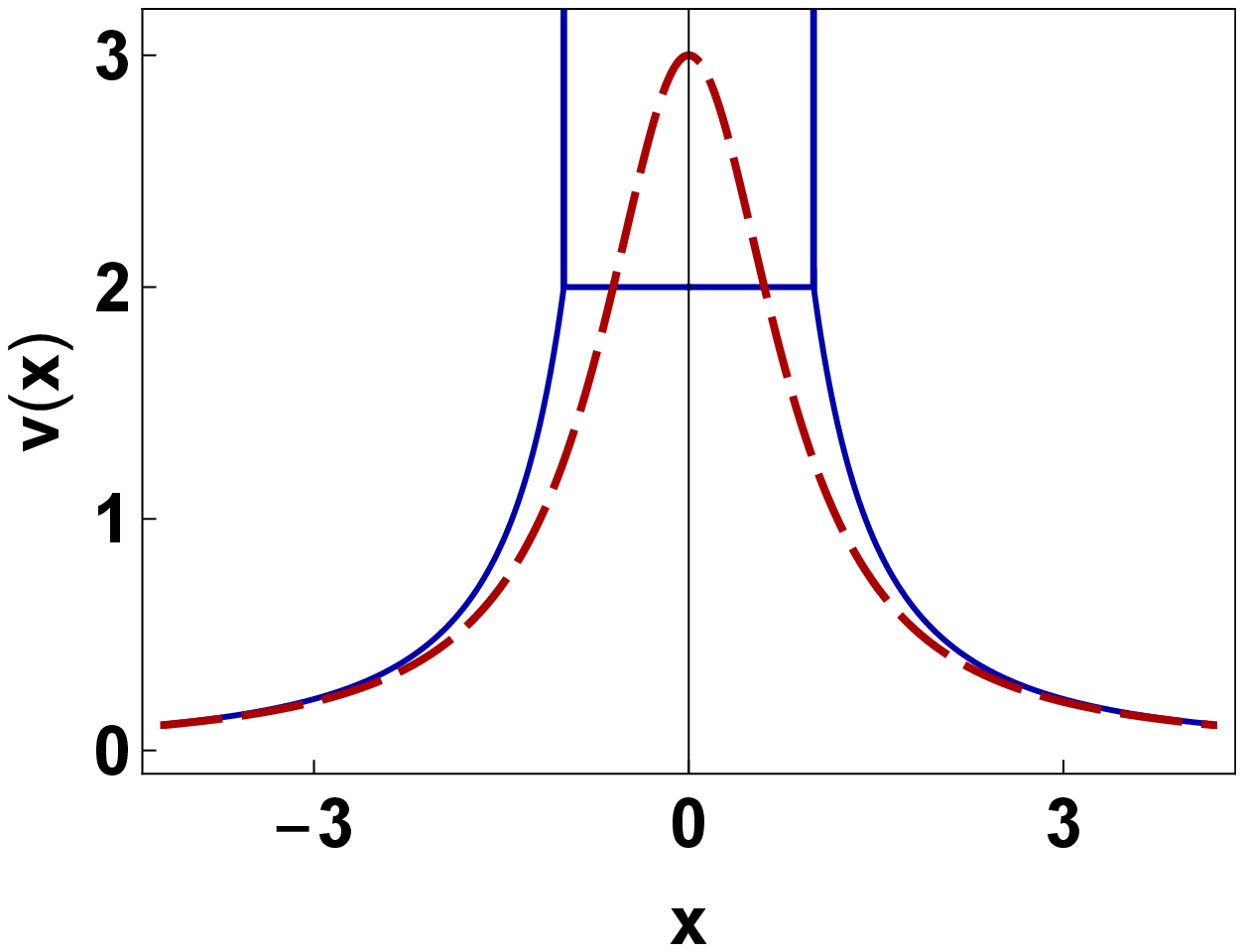}
  	\end{center}
	\vspace{-12pt}
  	\caption{Plots of the potentials (\ref{v-Ellis}) for the Ellis wormhole (dashed red curve)
	and (\ref{v-st}) for a wormhole with sharp transition (solid blue curve) with $r_0=x_0=1$ and $l=0$ (on the left) and $l=1$ (on the right). The vertical lines represent the delta-functions appearing in (\ref{v-st}).}
	\label{fig2}
	\end{figure}

If $p(r)=1$, (\ref{potential}) takes the following simple form.
    \be
    v(x):=\frac{l(l+1)}{r(x)^2}+\frac{r''(x)}{r(x)}.
    \label{potential-p1}
    \ee
For the Ellis wormhole $r(x)$ is given by (\ref{ellis-wh}), and (\ref{potential-p1}) yields
	\be
	v(x)=v_\tE(x):=\frac{l(l+1)}{x^2+r_0^2}+\frac{r_0^2}{(x^2+r_0^2)^2}.
	\label{v-Ellis}
	\ee
For a wormhole with a sharp transition, we can use (\ref{sharp}) to show that
	\bea
	r(x)&:=&r_0+(|x|-x_0)\theta(|x|-x_0)\nn\\
	&=& r_0+(x-x_0)\theta(x)\theta(x-x_0)-(x+x_0)\theta(-x)\theta(-x-x_0),
	\label{r0=}
	\eea
where $\theta(x)$ stands for the Heaviside step function. Substituting this equation in (\ref{potential-p1}), we find the following expression for the scattering potential associated with the sharp-transition wormhole.
	\be
	v(x)=v_\tst(x):=\frac{l(l+1)w(x)}{r_0^2}+\frac{1}{r_0}\left[\delta(x+x_0)+\delta(x-x_0)\right],
	\label{v-st}
	\ee
where
	\be	
	w(x):=\left\{\begin{array}{ccc}
	1 & {\rm for} & |x|\leq x_0,\\[3pt]
	\displaystyle \frac{r_0^2}{(|x|-x_0+r_0)^2} & {\rm for} & |x|> x_0.\end{array}\right.\nn
	\ee
Fig.~\ref{fig2} shows the plots of these potentials for $r_0=x_0=1$, ${l}=0$ and $l=1$.
	
In general, for $x\to\pm\infty$, $r\to |x|$ and $p(r)\to 1$, in view of (\ref{u-def}) this implies that the potentials (\ref{potential}) and (\ref{potential-p1}) have identical asymptotic behavior. The first term on the right-hand side of (\ref{potential-p1}) is a short range potential.\footnote{A potential is said to be a short-range potential provided that, for $|x|\to\infty$, it decays to zero faster than a constant multiple of the $1/|x|$. This is certainly the case for the potentials (\ref{v-Ellis}) and (\ref{v-st}).} The same holds for the second term provided that there is a positive real number $\alpha$ such that
	\be
	\lim_{x\to\pm\infty} x^\alpha r''(x)= 0.
	\label{short-range}
	\ee
If this condition holds, (\ref{potential}) is a short-range potential and we can employ the standard tools of potential scattering for short-range potentials \cite{reed-simon} to address the scattering problem for the scalar fields by the wormhole. 

We close this section by the following remarks.
\begin{enumerate}
\item For $l=0$, $v_\tst$ coincides with the double-delta function potential,
	\be
	v_\tst(x):=\frac{1}{r_0}\left[\delta(x+x_0)+\delta(x-x_0)\right],
	\label{v-st-zero}
	\ee
whose scattering properties are quite well-known \cite{reading-1972,Hakke-1981,Besprosvany-2001,jpa-2009,pra-2012,springer-2018}.
\item For $l>0$, we can identify the term $l(l+1)/r(x)$ in the potentials (\ref{potential}) and (\ref{potential-p1}) with a barrier potential whose hight is proportional to $l(l+1)$. As a low-energy wave passes through such a barrier, it is exponentially attenuated. Therefore, the contribution of this term to the transmission coefficient of the potentials (\ref{potential}) and (\ref{potential-p1}) is negligibly small. In other words, for low-energy incident waves only their s-wave component contributes to the transmission through the wormhole. Because the traversable wormholes of our interest are fragile, they can serve as the background spacetime only for the low-energy waves. The above argument shows that if the source of such a low-energy incident wave is located in one of the universes connected through the wormhole,  the non-spherical components of the wave gets filtered by the wormhole as it passes through its throat, and the transmitted wave that reaches an observer $O^-$ residing in the other universe will be spherical; such a wormhole will then appear to $O^-$ as a source of low-energy s-waves. An observer $O^+$ residing in the same universe as the source of the incident wave will see part of the spherical components of this wave gets absorbed. The details of the frequency-dependence of the corresponding absorption cross section can then provide crucial information for $O^+$ to decide if the source of the absorption is a wormhole or some other celestial body such as a blackhole. 

\item In view of the preceding remark, the transmission of low-energy waves through a  wormhole with $p(r)=1$ is governed by the potential,
	\be
	v(x):=\frac{r''(x)}{r(x)}.
	\label{v-main}
	\ee
Because the wormhole is asymptotically flat, we demand that $r(x)$ admits the asymptotic expression,
	\be
	r(x)=\pm x +c_\pm+\frac{f_{\pm}(x^{-1})}{x^{\epsilon}}~~~{\rm as}~~~x\to\pm\infty,
	\label{r-L12}
	\ee
where $c_\pm$ and $\epsilon$ are real constants, $\epsilon>0$, and $f^\pm:\R\to\R$ are functions that are analytic at $0$. Using (\ref{r-L12}), we can easily verify that $r''(x)$ satisfies (\ref{short-range}) for $\alpha= 2$, and (\ref{v-main}) defines a short-range potential belonging to the function space $L^1_2(\R)$, where $L^1_\sigma(\R)$ with $\sigma\in\R$ stands for the space of functions $\xi:\R\to\C$ fulfilling $\int_{-\infty}^\infty dx\:(1+|x|^\sigma)|\xi(x)|<\infty$. 

\item For the cases where $p(x)\neq 1$, (\ref{potential}) is an energy-dependent potential. In this sense, it is analogous to the optical potentials one encounters in the study of the propagation of light in effectively one-dimensional dielectric media. This is turn suggests the possibility of realizing optical analogs of wormhole scattering.
\end{enumerate}

\section{Potential scattering and its dynamical formulation}

The solutions of the Schr\"odinger equation (\ref{sch-eq}) for a short-range potential $v(x)$ and $k\in\R^+$ have plane-wave asymptotics, i.e., there are possibly $k$-dependent complex coefficients $A_\pm$ and $B_\pm$ such that
	\be
	\psi(x)\to A_\pm\, e^{ikx}+B_\pm\, e^{-ikx}~~~{\rm for}~~~x\to\pm\infty.
	\label{asymp}
	\ee
The transfer matrix for such a potential is the $k$-dependent $2\times 2$ matrix $\bM$ that relates $A_\pm$ and $B_\pm$ according to
	\be
	\left[\begin{array}{c}
	A_+\\
	B_+\end{array}\right]=\bM
	\left[\begin{array}{c}
	A_-\\
	B_-\end{array}\right].
	\label{M-def}
	\ee
This relation determines $\bM$ in a unique manner, if we demand that it is independent of $A_-$ and $B_-$, \cite{epjp-2019}.

Among the solutions of the Schr\"odinger equation (\ref{sch-eq}) there are the so-called scattering or Jost solutions $\psi_{l/r}$ that, by definition, satisfy
	\be
	\begin{aligned}
	&\psi_{l}(x)\to\left\{\begin{array}{ccc}
	e^{ikx}+R^l e^{-ikx} & {\rm for} & x\to-\infty,\\
	T e^{ikx}  & {\rm for} & x\to\infty,\end{array}\right.\\
	&\psi_{r}(x)\to\left\{\begin{array}{ccc}
	T e^{-ikx} & {\rm for} & x\to-\infty,\\
	e^{-ikx}+R^r e^{ikx}  & {\rm for} & x\to\infty,\end{array}\right.
	\end{aligned}
	\label{jost}
	\ee
where $R^{l/r}$ and $T$ are the left/right reflection and transmission amplitudes of the potential. In view of (\ref{asymp}) -- (\ref{jost}), we can express them in terms of the entries $M_{ij}$ of $\bM$;
	\begin{align}
	&R^l=-\frac{M_{21}}{M_{22}},
	&& R^r=\frac{M_{12}}{M_{22}},
	&& T=\frac{1}{M_{22}}.
	\label{RRT}
	\end{align}
These in turn imply
	\be
	\bM=\frac{1}{T}\left[\begin{array}{cc}
	T^2-R^lR^r & R^r\\
	-R^l & 1\end{array}\right].
	\label{M=}
	\ee
The solution of the scattering problem for a short-range potential $v(x)$ means the determination of its reflection and transmission amplitudes. According to (\ref{M=}), this is equivalent to finding the transfer matrix of the potential.

The transfer matrix has been used as an effective tool for the study of wave propagation in multilayer media since the 1940's  \cite{jones-1941,abeles,thompson,yeh}. This is mainly because of its composition property which allows for expressing the transfer matrix of such a medium as the product of the transfer matrices for its layers. In terms of the scattering potential $v(x)$ modeling the medium, this means that if we dissect $\R$ into $N$ intervals $(a_{i-1},a_{i})$, with $i\in\{1,2,\cdots,N\}$ and
	\[-\infty=:a_0<a_1<a_2<\cdots<a_{N-1}<a_N:=\infty,\]
and let $\bM_i$ be the transfer matrix for the truncated potential,
	\[ v_i(x):=\left\{\begin{array}{ccc}
	v(x) &{\rm for}& x\in (a_{i-1},a_{i}),\\
	0 &{\rm for}& x\notin (a_{i-1},a_{i}),\end{array}\right.\]
then the transfer matrix of $v(x)$ is given by \cite{springer-2018}
	\be
	\bM=\bM_N\bM_{N-1}\cdots\bM_1.
	\label{compose}
	\ee
Refs.~\cite{ap-2014,pra-2014a} offer a straightforward derivation of this relation which is based on a curious identification of the transfer matrix with the $S$-matrix of an effective non-unitary two-level quantum system.

Consider a two-level system whose dynamics is determined by the time-dependent Schr\"odinger equation, $i\partial_x\Psi(x)=\bH(x)\Psi(x)$, where $x$ plays the role of time, $\bH(x)$ is the interaction-picture Hamiltonian given by,
	\be
	\bH(x):=\frac{v(x)}{2k}\left[\begin{array}{cc}
	1 & e^{-2ikx}\\
	-e^{2ikx} & -1\end{array}\right],
	\label{H=1}
	\ee
and $v(x)$ is a given short-range scattering potential. Let $\bU(x,x_0)$ denote the evolution operator defined by
	\be
	i\partial_x\bU(x,x_0)=\bH(x)\bU(x,x_0),~~~~~~~\bU(x,x_0)=\bI,
	\label{sch-eq-U}
	\ee
where $x_0$ is an initial value of $x$, and $\bI$ is the $2\times 2$ identity matrix. We can express $\bU(x,x_0)$ as the time-ordered exponential,
	$\bU(x,x_0)=\sT\exp\left\{-i\int_{x_0}^x dx'\ \bH(x') \right\}$,
where $\sT$ is the time-ordering operator. Ref.~\cite{ap-2014} shows that
	\bea
	\bM&=&\bU(\infty,-\infty)=\sT\exp\left\{-i\int_{-\infty}^\infty dx\ \bH(x) \right\}
	\label{M=Texp-1}\\
	&=&\bI+\sum_{\ell=1}^\infty(-i)^\ell
    \int_{-\infty}^\infty dx_\ell \int_{-\infty}^{x_\ell}dx_{\ell-1}\cdots
    \int_{-\infty}^{x_{2}}dx_{1}\bH(x_\ell)\bH(x_{\ell-1})\cdots\bH(x_1).
	\label{M=Texp}
	\eea
This result has found interesting applications in addressing a number of basic problems of scattering theory \cite{jpa-2014a,jpa-2014b,pra-2014b,pra-2016,pra-2017,jpa-2018,jmp-2019,jpa-2020a,jpa-2020b}.

A simple application of (\ref{M=Texp}) is in treating delta-function potentials,
	\be
	v_a(x):=\fz\,\delta(x-a),
	\label{delta-1}
	\ee
where $a\in\R$ is arbitrary. To see this first we express (\ref{H=1}) in the form,
	 \be
	\bH(x):=\frac{v(x)}{2k}\, e^{-ikx\bsigma_3}\bcK \, e^{ikx\bsigma_3},
	\label{H=2}
	\ee
where $\bsigma_j$ with $j\in\{1,2,3\}$ are Pauli matrices, and $\bcK:=\bsigma_3+i\bsigma_2$.
We then substitute (\ref{delta-1}) in (\ref{H=2}) and use the identity, $\bcK^2=\bzero$, to establish,
	\begin{align}
	&\bH(x_1):=\frac{\fz}{2k}e^{-ika\bsigma_3}\bcK\, e^{ika\bsigma_3}\delta(x_1-a),
	&&\bH(x_2)\bH(x_1)=\bzero,
	\label{H=delta}
	\end{align}
where $x_1,x_2\in\R$ are arbitrary and $\bzero$ labels the $2\times 2$ null matrix. In view of the second equation in (\ref{H=delta}), the Dyson series in (\ref{M=Texp}) truncates, and we obtain the following formula for the transfer matrix of the delta-function potential $v_a$.
	\be
	\bM=\bM_a:=\bI-\frac{i\fz}{2k}e^{-ika\bsigma_3}\bcK\, e^{ika\bsigma_3}=
	\frac{1}{2k}\left[\begin{array}{cc}
	2k-i\fz & -i\fz e^{-2ika}\\
	i\fz e^{2ika} & 2k+i\fz
	\end{array}\right].
	\label{M=delta}
	\ee

\section{Scattering by a wormhole with sharp transition}

For a wormhole with a sharp transition, $p(r)=1$, and the separable solutions (\ref{basic-sols}) of the field equation (\ref{scalar-field}) take the form,
	\be
	\Phi(t,x,\vartheta,\varphi)=e^{-i\omega t}{Y_l^m}(\vartheta,\varphi)\,\frac{\psi(x)}{r(x)},
	\label{sep-sol-p=1}
	\ee
where $r(x)$ is given by (\ref{sharp}) and $\psi(x)$ is a solution of the Schr\"odinger equation (\ref{sch-eq}) for the potential (\ref{v-st}). This equation admits exact scattering solutions in terms of the spherical Hankel functions. Therefore, the scattering problem for this wormhole is exactly solvable for all angular momentum quantum numbers $l$. Because the formulas for the reflection and transmission amplitudes are rather complicated and we are mainly interested in low-energy waves that can pass through the wormhole, here we confine our attention to the $s$-waves, i.e., $l=0$. We present the results for general $l$ in Appendix~A. 

As we noted above, for $l=0$, (\ref{v-st}) reduces to the double-delta-function potential (\ref{v-st-zero}). We can easily compute the transfer matrix of this potential using the formula (\ref{M=delta}) for the transfer matrix of the single-delta-function potential (\ref{delta-1}) and the composition rule (\ref{compose}). This yields
	\be
	\bM=\bM_{x_0}\bM_{-x_0}=
	\left[\begin{array}{cc}
	1-\frac{i}{\fK}\left\{1+\frac{e^{-2i\hx\fK}\sin(2\hx\fK)}{2\fK}\right\}
	&-\frac{i}{\fK}\left\{\cos(2\hx\fK)+\frac{\sin(2\hx\fK)}{2\fK}\right\}\\[6pt]
	\frac{i}{\fK}\left\{\cos(2\hx\fK)+\frac{\sin(2\hx\fK)}{2\fK}\right\}
	&1+\frac{i}{\fK}\left\{1+\frac{e^{2i\hx\fK}\sin(2\hx\fK)}{2\fK}\right\}\end{array}\right],
	\label{M-2delta}
	\ee
where $\bM_{\pm x_0}$ stands for $\bM_a$ of Eq.~(\ref{M=delta}) with $a=\pm x_0$ and $\fz=r_0^{-1}$, and we have introduced
	\begin{align*}
	&\fK:=kr_0, &&\hx:=\frac{x_0}{r_0}.
	\end{align*}
Expanding the right-hand side of (\ref{M-2delta}) in powers of $\fK$, we find
	\bea
	\bM
	&=&-\frac{i(1+\hx)}{\fK}\,\bcK+(1-2\hx^2)\bI+\frac{2}{3}\,\hx^2\,\fK\left[4i\hx\,\bsigma_3-
	(\hx+3)\bsigma_2\right]+{O}(\fK^2),
	\label{M-2delta-LE}
	\eea
where ${O}(\fK^n)$ stands for terms of order $n$ and higher in powers of $\fK$.

\begin{figure}
  	\begin{center}
 	\includegraphics[scale=0.55]{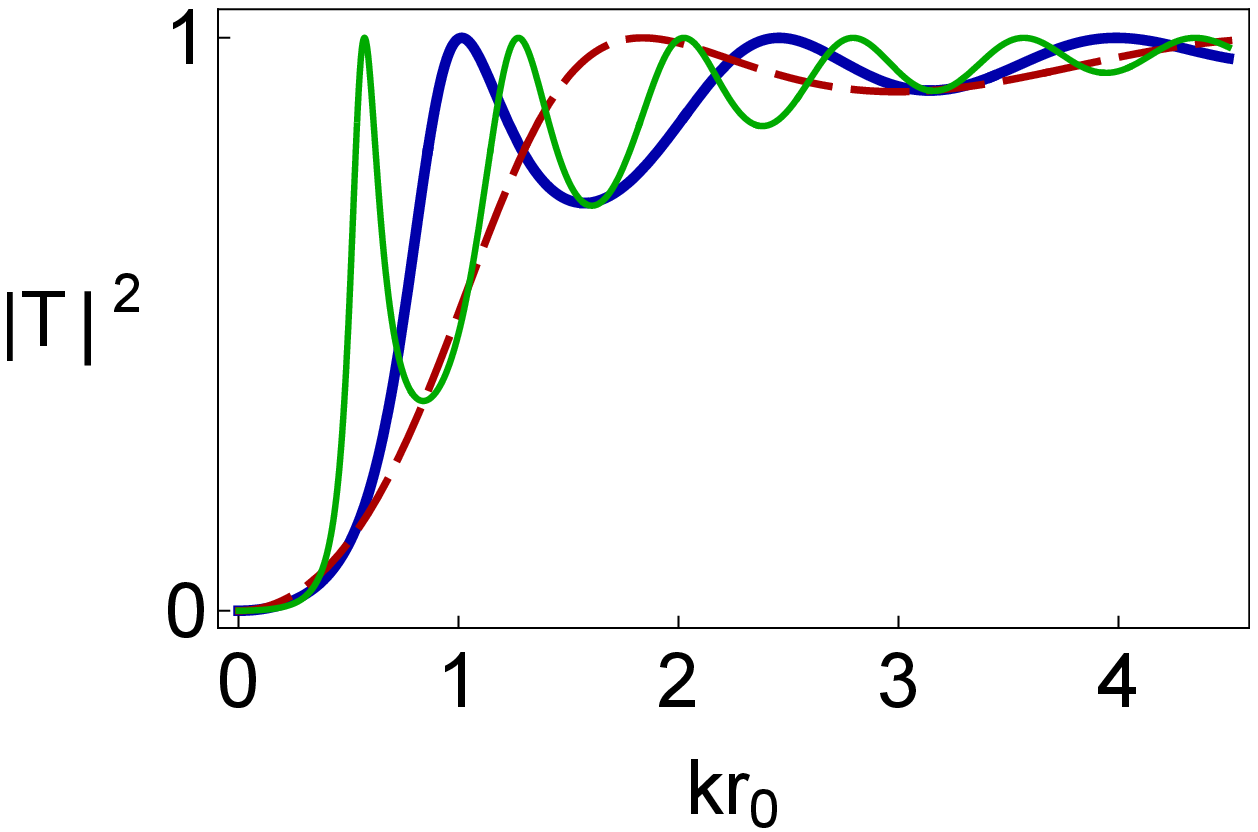}\hspace{1cm}
	\includegraphics[scale=0.55]{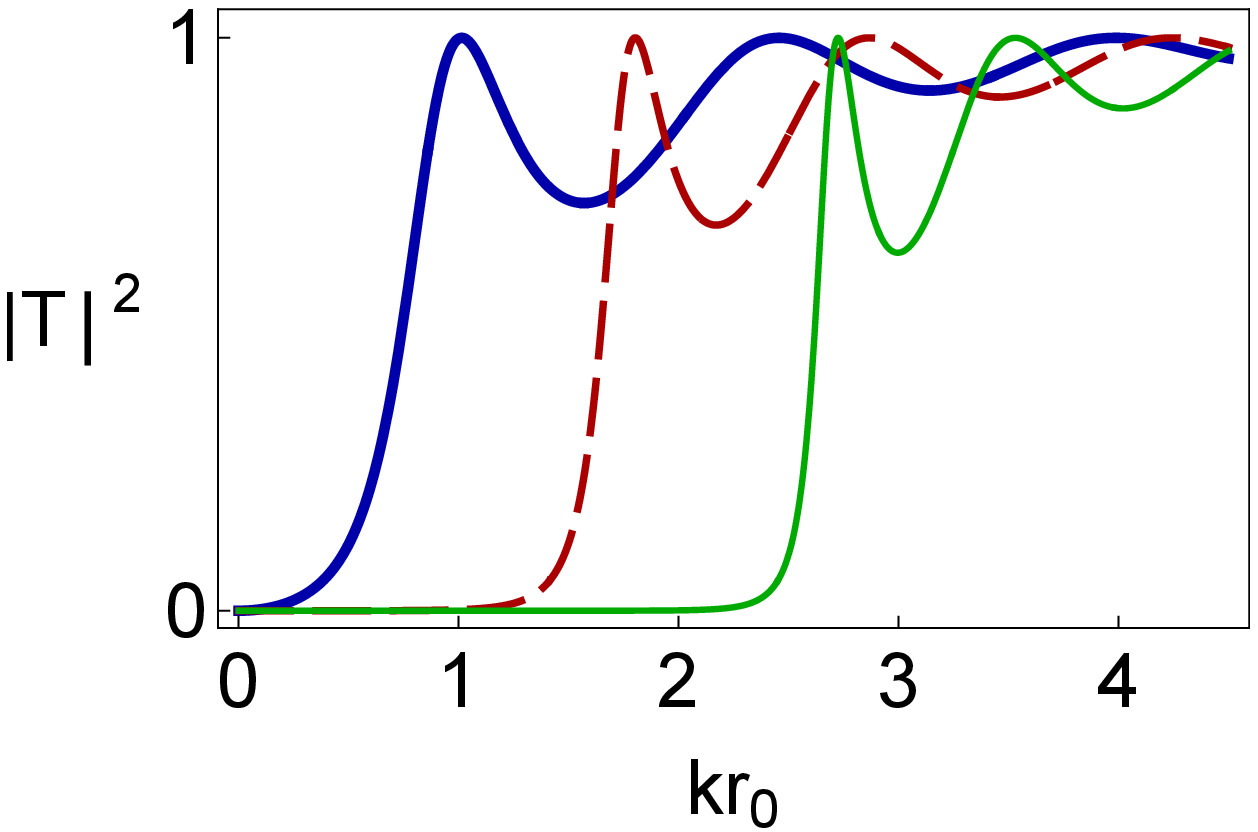}
  	\end{center}
  	\vspace{-12pt}
  	\caption{Graphs of the transmission coefficient $|T|^2$ as a function of $kr_0$ for a sharp-transition wormhole with $l=0$ and $x_0=r_0$ (thick solid blue curve), $x_0=r_0/2$ (dashed red curve), and $x_0=2r_0$ (thin solid green curve) on the left, and $x_0=r_0$ and $l=0$ (thick solid blue curve), $l=1$ (dashed red curve), and $l=2$ (thin solid green curve) on the right.}
	\label{fig3}
	\end{figure}

In view of (\ref{RRT}), (\ref{M-2delta}), and (\ref{M-2delta-LE}), we find the following expression for the reflection and transmission amplitudes of the sharp-transition wormhole for s-waves.
	\bea
	&&R^l=R^r=
	\frac{-i[\cos(2\hx\fK)+\sin(2\hx\fK)/2\fK]}{\fK+i[1+e^{2i\hx\fK}\sin(2\hx\fK)/2\fK]}=
	-1+\frac{i(2\hx^2-1)\fK}{\hx+1}+{O}(\fK^2),
	\label{R-2delta-LE}
	\\
	&&T=\frac{\fK}{\fK+i[1+e^{2i\hx\fK}\sin(2\hx\fK)/2\fK]}
	=-\frac{i\fK}{\hx+1}-\frac{(2\hx^2-1)\fK^2}{(\hx+1)^2}+{O}(\fK^3).
	\label{T-2delta-LE}
	\eea
In Appendix~A, we give the generalization of these equations for the scalar waves with arbitrary $l$. For $\fK\ll 1$, the leading order contribution to $T$ turns out to be proportional to $\fK^{2l+1}$. This confirms our expectation that the transmission coefficient $|T|^2$ for low-energy waves with $l>0$ is much smaller than the one for $l=0$. Fig.~\ref{fig3} provides a graphical demonstration of the behavior of $|T|^2$ for different values of $\hx$ and $l$.

\section{Low-energy scattering by a wormhole with $p(r)=1$}
	
For $p(r)=1$, Eq.~(\ref{WH}) corresponds to a class of ultrastatic wormholes that includes the Ellis wormhole \cite{sonego-2010,Tsukamoto}. For this class of wormholes Eq.~(\ref{sep-sol-p=1}) still holds, but we do not have access to the exact scattering solutions of the Schr\"odinger equation~(\ref{sch-eq}). We can still use the general results on low-energy potential scattering to determine the structure of the reflection and transmission amplitudes for small values of the wavenumber $k$. These results which have been obtained within the framework of mathematical scattering theory apply to specific classes of potentials. For example, Refs.~\cite{newton-1986,aktosun-2001} report the basic results on low-energy scattering for potentials belonging to $L^1_\sigma(\R)$ with $1\leq\sigma\leq 2$. As we noted in Sec.~2, for the class of wormholes with $p(r)=1$ and $r(x)$ having the asymptotic form (\ref{r-L12}), the problem of the transmission of low-energy scalar waves is described by the potential (\ref{v-main}) which belongs to $L^1_2(\R)$. Therefore, we can benefit from the results reported in \cite{newton-1986,aktosun-2001}. Recently, two of us have obtained an alternative prescription for determining the low-energy scattering properties of this class of potentials \cite{p162}. In the following, we describe this prescription and discuss its application for the scattering of low-energy scalar waves passing through a wormhole with $p(r)=1$.

\subsection{Transfer matrix at low energies}

The transfer matrix $\bM$ for a scattering potential $v(x)$ belonging to $L^1_2(\R)$ admits a series expansion of the form,
	\be
	\bM=\frac{1}{k}\ \bm^{(-1)}+\bm^{(0)}+{o}(k^0),
	\label{M-expand}
	\ee
where $\bm^{(n)}$ are $k$-independent matrices and ${o}(k^\alpha)$ stands for a function of $k$ such that ${o}(k^\alpha)/k^\alpha\to0$ as $k\to 0$.  A key observation made in Ref.~\cite{p162} is that the evolution operator $\bU(x,x_0)$, which gives $\bM$ via (\ref{M=Texp-1}), satisfies 
	\be
	\bU(x,-\infty)=-\frac{i}{2k}\, \phi'_1(x)\bcK+ \frac{1}{2}\Big\{
	[\phi_1(x)-x\phi'_1(x)](\bI+\bsigma_1)+\phi'_2(x)(\bI-\bsigma_1)\Big\}+o(k^0),
	\label{M1-9}
	\ee
where $\phi_1,\phi_2:\R\to\C$ solve the zero-energy Schr\"odinger equation,
	\be
	-\phi''(x)+v(x)\phi(x)=0.
	\label{sch-eq-0}
	\ee
Demanding that $\bU(-\infty,-\infty)=\bI$, we find 
	\be
	\begin{aligned}
	&\lim_{x\to-\infty}[\phi_1(x)-x\phi_1'(x)]=1,\quad\quad\quad\quad\lim_{x\to-\infty}\phi'_1(x)=0,\\
	&\lim_{x\to-\infty}[\phi_2(x)-x\phi_2'(x)]=0,\quad\quad\quad\quad\lim_{x\to-\infty}\phi'_2(x)=1.
	\end{aligned} 
	\label{IC-1}
	\ee
Substituting (\ref{M1-9}) in (\ref{M=Texp-1}), we obtain (\ref{M-expand}) with 
	\begin{align}
	&\bm^{(-1)}=-\frac{i\fa_1}{2}\,\bcK,
	&&\bm^{(0)}=\frac{1}{2}[{\fb}(\bI+\bsigma_1)+\fa_2(\bI-\bsigma_1)],
	\label{order-0m1}
	\end{align}
where, for $i\in\{1,2\}$,
	\begin{align}
	&\fa_i:=\lim_{x\to+\infty}\phi'_i(x),
	&&\fb:=\lim_{x\to+\infty}[\phi_1(x)-x\phi_1'(x)].
	\label{abs-def}
	\end{align}

Next, we note that for every potential $v\in L^1_2(\R)$, there is a solution $\phi_0:\R\to\C$ of (\ref{sch-eq-0}) and real numbers $c_\pm$ such that as $x\to\pm\infty$
	\begin{align}
	&\phi_0(x)=\pm x+c_\pm+o(x^0),
	&&\phi_0'(x)=\pm 1+o(x^{-1}).
	\label{condi-0}
	\end{align}
We can use these relations to show that
	 \begin{align}
    	&\phi_1(x)= \phi_0(x) \int_{-\infty}^x \frac{du}{\phi_0(u)^2}, 
    	&&\phi_2(x)=c_-\phi_1(x)-\phi_0(x).
	\label{phi12=}
    	\end{align}
Inserting these in (\ref{abs-def}), we find
	\begin{align}
	&\fa_1={s}:=\int_{-\infty}^\infty \frac{dx}{\phi_0(x)^2},
	&&\fa_2=c_-{s}-1,&&{\fb}=c_+{s}-1.
	\label{ab=2}
	\end{align}		
In view of (\ref{M-expand}) and (\ref{order-0m1}) -- (\ref{ab=2}), 
	\be
    	\bM=-\frac{i{s}}{2k}\,\bcK+
	\frac{1}{2}\Big\{\left[(c_++c_-){s}-2\right]\bI+
	(c_+-c_-){s}\, \bsigma_1\Big\}+o(k^0).
	\label{M=01-order}
	\ee

Finally, we use (\ref{RRT}) and (\ref{M=01-order}) to derive the following asymptotic expressions for the reflection and transmission amplitudes.
	\bea
	R^l&=&-1-2i\left(c_--{s}^{-1}\right)k+o(k^1),
	\label{RL=}\\
	R^r&=&-1-2i\left(c_+-{s}^{-1}\right)k+o(k^1),
	\label{RR=}\\
	T &=&- 2i{s}^{-1}k+2{s}^{-1}(c_++c_--2{s}^{-1})k^2 +o(k^2).
	\label{T=}
	\eea
If $v$ is an even function, we can use the fact that $\phi_0$ satisfies (\ref{sch-eq-0}) to show that $\phi_0$ is also an even function. In light of (\ref{phi12=}), this implies that $c_-=c_+$, and (\ref{RL=}) -- (\ref{T=}) reduce to
	\bea
	&&R^l=R^r=-1-2i\left(c_+-{s}^{-1}\right)k+o(k^1),
	\label{R=even}\\
	&&T=- 2i{s}^{-1}k+4{s}^{-1}(c_+-{s}^{-1})k^2 +o(k^2).
	\label{T=even}
	\eea

\subsection{Application to ultrastatic wormholes}

Consider a ultrastatic wormhole with $p(r)=1$ and $r(x)$ fulfilling (\ref{r-L12}). The low-energy scattering of scalar waves passing through such a wormhole is described by a potential of the form (\ref{v-main}) which belongs to $L^1_2(\R)$. The principal example is the Ellis wormhole and the family of its generalization corresponding to 
	\be
	r(x)=(x^{2n}+r_0^{2n})^{1/2n},
	\label{gen-Ellis}
	\ee
where $n\in\Z^+$, \cite{kar-1995}. In general, the function $r(x)$ has a minimum value, corresponding to the radius of the wormhole's throat, which is attained at $x=0$. In analogy to (\ref{gen-Ellis}), we denote it by $r_0$, i.e., $r_0:=r(0)$.

In the preceding subsection we have reduced the problem of determining the low-energy asymptotic behavior of the reflection and transmission amplitudes of a potential belonging to the class $L^1_2(\R)$ to that of finding a solution $\phi_0$ of the zero-energy Schr\"odinger equation (\ref{sch-eq-0}) satisfying (\ref{condi-0}). For the potentials of the form (\ref{v-main}) that arise in the discussion of wormholes, the identification of $\phi_0$ requires no effort; according to (\ref{v-main}) and (\ref{r-L12}), the function $r(x)$ satisfies both (\ref{sch-eq-0}) and (\ref{condi-0}). Therefore, we only need to set $\phi_0(x)=r(x)$ for the calculation of the $c_\pm$ and ${s}$ that appears in (\ref{RL=}) -- (\ref{T=even}), i.e., substitute 
	\begin{align}
	&c_\pm=\lim_{x\to\pm\infty}[r(x)-|x|],
	&&{s}=\int_{-\infty}^\infty \frac{dx}{r(x)^2},
	\label{c-s=}
	\end{align}
in these equations. 

For a wormhole with a sharp transition that is defined using (\ref{sharp}), $v$ is the double-delta-function potential (\ref{v-st-zero}) which is even, and (\ref{c-s=}) gives
	\begin{align}
	&c_\pm=r_0-x_0=r_0(1-\hx),
	&&{s}=\frac{2(x_0+r_0)}{r_0^2}=\frac{2(\hx+1)}{r_0},
	\end{align} 
where $\hx:=x_0/r_0$. Plugging these equations in (\ref{R=even}) and (\ref{T=even}), we recover (\ref{R-2delta-LE}) and (\ref{T-2delta-LE}).

For the generalized Ellis wormholes corresponding to (\ref{gen-Ellis}),  $v$ is again an even potential, and (\ref{c-s=}) yields
	\begin{align*}
	&c_\pm=0, &&s=\frac{2^{2-1/n}\sqrt{\pi}\ 
	\Gamma(1+\frac{1}{2n})}{r_0\Gamma(\frac{1+n}{2n})},
	\end{align*} 
where $\Gamma$ denoting the Gamma function, and (\ref{R=even}) and (\ref{T=even}) become
	\begin{align}
	&R^l=R^r=-1+\frac{2i k}{s} +{o}(k^1),
	&&T=-\frac{2ik}{s}-\frac{4k^2}{s^2}+{o}(k^2).\nn
	\end{align}
For the Ellis wormhole, which corresponds to setting $n=1$, $s=\pi/ r_0$.

\section{Discussion and conclusions}

In potential scattering in one dimension, both the source of the wave and the detector can be placed at $x=\pm\infty$. Given a source located at $x=+\infty$ the intensities of the wave detected by an observer at $x=+\infty$ and $x=-\infty$ are respectively proportional to the reflection and transmission coefficients, $|R^r|^2$ and $|T|^2$. The same holds for the scattering of a scalar wave by a wormhole. If the source of the incident wave lies in the region of the space corresponding to $x=+\infty$, an observe $O^+$ located in the same region will detect the scattered wave that reaches $x=+\infty$, while an observe $O^-$ located in the other side of the wormhole, at $x=-\infty$, will detect the transmitted wave. 

For the observer $O^+$ the transmission of the wave through the wormhole will appear as the partial absorption of the spherical component of the wave by a celestial body. As we show in Appendix~B, the absorption cross section has the form $\sigma=\pi|T|^2/k^2$. Substituting (\ref{T=}) in this equation and making use of (\ref{c-s=}), we find
	\be
	\sigma=A+{o}(k),
	\label{sigma=}
	\ee
where $A:=4\pi \rho^2$ and $\rho:=1/s$. Notice that $o(k)$ on the right-hand side of (\ref{sigma=}) implies that the coefficient of linear term in $k$ vanishes. 

It is well-known that for a spherically symmetric blackhole the zero-frequency limit of the absorption cross section associated with a minimally coupled massless scalar field coincides with the area of its event horizon \cite{das-1997}. See also \cite{Higuchi-2001,Magalhes-2020}. Eq.~(\ref{sigma=}) shows that using the extreme low-energy scattering data for a massless field the observer $O^+$ cannot distinguish between the wormhole and a blackhole with event horizon area $A$ (radius $\rho$.) 

Ref.~\cite{Unruh-1976} calculates the same absorption cross section for an arbitrary scalar field that is minimally couple to a Schwarzschild blackhole. In our notation, the result of this calculation, i.e., Eq.~(78) of \cite{Unruh-1976}, reads
	\be
	\sigma=\frac{\sqrt{\pi} A^{3/2}(v^2+1)\omega}{
	2v^2[1-e^{-\sqrt{\pi A}(v+v^{-1})\omega/2}]},
	\label{unruh}
	\ee
where again $A$ is the area of the event horizon, and $\omega$ and $v:=k/\omega$ are respectively the angular frequency and speed of the incident wave. For a massless field, where $\omega=k$ and $v=1$, this formula reduces to
	\be
	\sigma=\frac{\sqrt{\pi} A^{3/2}\,\omega}{
	1-e^{-\sqrt{\pi A}\, \omega}}=A+\frac{\sqrt{\pi}}{2}\,A^{3/2} k+
	{O}(k^2).
	\label{unruh-0}
	\ee
Comparing (\ref{sigma=}) and (\ref{unruh-0}), we see that unlike the case of a wormhole, the slop of the absorption cross section, $\partial_k\sigma$, for Schwarzschild blackhole tends to a nonzero value as $k\to 0$. Therefore, the observer $O^+$ can distinguish the wormhole from a Schwarzschild blackhole, if her low-energy scattering data allows her to decide about the next-to-leading order contribution to $\sigma$ in powers of $k$.  

The situation is different for a massive field. In the limit that $k\to 0$, $\omega\to m$, $v\approx k/m$, and (\ref{unruh}) gives $\sigma\approx \sqrt{\pi}\, m^3 A^{3/2}/2k^2 $, i.e., the blackhole's absorption cross section diverges quadratically as $k\to 0$. This is in sharp contrast to the case of a wormhole, for according to (\ref{sigma=}), in this limit the absorption cross section of the wormhole for a massive scalar field tends to $A$ which is finite. Therefore, the observer $O^+$ can easily differentiate between a wormhole and a Schwarzschild blackhole by examining the scattering data for a massive scalar field whose wavenumber is much smaller than its mass; $k\ll m$.

 The observer $O^-$ will view the wormhole as a source of spherical waves whose intensity is proportional to the cross section (\ref{sigma=}). In the limit $k\to 0$, the wormhole will appear to $O^-$ as a whitehole, if she detects low-energy massless scalar fields. 
 
Our results apply to wormholes determined by the line element (\ref{WH}) with $p(r)=1$ and $r(x)$ having the asymptotic form (\ref{r-L12}). They are obtained under the assumption that $r_0k\ll 1$. Because $r_0$ signifies the size of the wormhole, these results hold for a wide range of frequencies provided that the wormhole is sufficiently small; for a wormhole with $r_0\approx 100~{\rm nm}$, they even apply to frequencies in the ultraviolet spectrum. If the wormhole is unstable, the observer $O^-$ will see it emit a spherical pulse and disappear.

\subsection*{Acknowledgements}
We are grateful to Behnaz Azad for her help in producing Fig.~\ref{fig1}. This project has been supported by Turkish Academy of Sciences (T\"UBA).

\section*{Appendix A: $R^{l/r}$ and $T$ for the potential~(\ref{v-st})}

In this appendix we outline the derivation of the formulas for the reflection and transmission amplitudes for the potential (\ref{v-st}). We can write this potential in the form $v=v_0+v_1+v_2$, where 	
	\bea
	v_0(x)&:=&\frac{1}{r_0}\left[\delta(x-x_0)+\delta(x+x_0)]\right],\nn\\
	v_1(x)&:=&\frac{l(l+1)}{r_0^2}\:\theta(x_0-|x|),\nn\\
	v_2(x)&:=&\frac{l(l+1)}{(|x|-x_0+r_0)^2}\:\theta(|x|-x_0).\nn
	\eea
To solve the time-independent Schr\"odinger equation (\ref{sch-eq}) for this potential it is sufficient to use the well-known solution of this equation for the barrier potential $v_1$ in the interval $[-x_0,x_0]$ and use the delta functions in $v_0$ to patch them to the solutions of (\ref{sch-eq}) for the truncated inverse square potential $v_2$ in $(-\infty,-x_0]$ and $[x_0,\infty)$. The latter are linear combinations of the functions, $g_i(y):=y\, h^{(i)}_l(y)$, where $i\in\{1,2\}$, $y:=k(|x|-x_0+r_0)$, and $h^{(i)}_l$ label  the spherical Hankel functions of order $l$.  Using the asymptotic expression for these functions we can determine the appropriate linear combinations of $g_i(y)$ that yield the Jost solutions (\ref{jost}) and read off the reflection and transmission amplitudes. This gives
	\begin{align}
	&R^l=R^r= X\left(Y+Z\right), &&T= X\left(Y-Z\right),
	\label{RT-app-A}
	\end{align}
where
	\begin{align*}
	& X:=\frac{(-1)^{(l+1)} e^{2i\fK(1-\hx)} g_2(\fK)}{2g_1(\fK)},
	&& Y:=\frac{W^*-\fK^{-1}(q\cot q\hx+1)}{W-\fK^{-1}(q\cot q\hx+1)},\\
	& Z:=\frac{W^*+\fK^{-1}(q\tan q\hx-1)}{W+\fK^{-1}(q\tan q\hx-1)},
	&& W:=\frac{g_1'(\fK)}{g_1(\fK)},
    	\end{align*}
$\fK:=r_0k$, $\hx:=x_0/r_0$, and $q:=\sqrt{\fK^2-l(l+1)}$.

\section*{Appendix B: Determination of the absorption cross section}

Consider the propagation of scalar fields in a wormhole background with $p(r)=1$. Recall that in our notation $x$ is the extended radial coordinate, which takes values in the whole real line, and that $x=\pm\infty$ mark spatial regions in the universes $M_\pm$ connected by the wormhole. The scattering setup where the source of the incident wave lies at $x=\infty$ corresponds to a solution, 
	\be
	\Phi(t,x,\vartheta,\varphi)=e^{-i\omega t}\Psi(x,\vartheta,\varphi),
	\label{eq-AppB1}
	\ee 
of the field equation (\ref{scalar-field}) such that $\Psi$ satisfies the following asymptotic boundary conditions.
	\be
	\Psi(x,\vartheta,\varphi)\to\left\{\begin{array}{ccc}
	e^{i\bk_0\cdot\bx}+f^+(\vartheta,\varphi)\: x^{-1}e^{ik x}
	&{\rm for}& x\to +\infty,\\[6pt]
	f^-(\vartheta,\varphi)\:(-x)^{-1} e^{-ik x}&{\rm for}& x\to-\infty,
	\end{array}\right.
	\label{f-pm}
	\ee
where 
$\bk_0$ is the incident wave vector, $\bx:=|x|(\sin\vartheta\cos\varphi,\sin\vartheta\sin\varphi,\cos\vartheta)$, and $f^\pm$ are the scattering amplitudes measured by observers $O^\pm$ residing at $x=\pm\infty$. Our aim is to relate $f^\pm$ to the reflection and transmission amplitudes of the potential (\ref{potential-p1}) associated with the wormholes with $p(r)=1$.  

First, we multiply both sides of (\ref{sep-sol-p=1}) by $Y_l^m(\vartheta,\varphi)^*$ and integrate over the unit sphere $S^2$. In view of the orthonormality of the spherical harmonics and Eq.~(\ref{eq-AppB1}), this gives
	\be
	\psi(x)=r(x)\int_{S^2}d\Omega\: Y_l^m(\vartheta,\varphi)^*\Psi(x,\vartheta,\varphi),
	\label{eq-AppB2}
	\ee
where $d\Omega:=\sin\vartheta d\vartheta d\varphi$. Substituting (\ref{f-pm}) in (\ref{eq-AppB2}) and making use of (\ref{r-L12}), we can show that
	\be
	\psi(x)\to \left\{
	\begin{array}{ccc}
	x\int_{S^2}d\Omega\:Y_l^m(\vartheta,\varphi)^*e^{i\bk_0\cdot\bx}+{f}^+_{l m}\:e^{ikx} &
	{\rm for}& x\to+\infty,\\[3pt]
	{f}^-_{l m}\:e^{-ikx}&{\rm for}& x\to-\infty,
	\end{array}\right.
	\label{eq-AppB3}
	\ee
where 
	\be
	{f}^\pm_{l m}:=\int_{S^2}d\Omega\:Y_l^m(\vartheta,\varphi)^*f^\pm(\vartheta,\varphi).
	\label{ff=}
	\ee
	
Next, we use the properties of the spherical harmonics \cite{sakurai-1997} to show that
 	\be
	\int_{S^2}d\Omega\: Y_l^m(\vartheta,\varphi)^* e^{i\bk_0\cdot\bx}=
	4\pi i^l\,\zeta_{lm}\, j_l(kx) ,
	\label{eq-AppB-id}
	\ee
where $j_l$ stands for the spherical Bessel functions, $\zeta_{lm}:=Y_m^l(\vartheta_0,\varphi_0)^*$, and $(\vartheta_0,\varphi_0)$ are the spherical angular coordinates marking the unit vector $\bk_0/k$, i.e., the direction of the incident wave. Eq.~(\ref{eq-AppB-id}) together with the fact that
	\[j_l(kx)\to \frac{i^{-(l+1)}}{2k x}\left[e^{ikx}-(-1)^l e^{-ikx}\right]~~~{\rm for}~~~x\to\infty,\]
allow us to express (\ref{eq-AppB3}) in the form
	\be
	\psi(x)\to 
	\frac{2\pi (-1)^l i \zeta_{lm}}{k}\,\times \left\{\begin{array}{ccc}	
	\displaystyle\left[e^{-ikx}+(-1)^{l+1}\left(1+\frac{i k{f}^+_{lm}}{2\pi \zeta_{lm}}\right)e^{ikx}
	\right] &{\rm for}& x\to+\infty,\\[9pt]
	\displaystyle\frac{(-1)^{l+1} ik{f}^-_{lm}}{2\pi \zeta_{lm}}e^{-ikx} &{\rm for}& x\to -\infty.
	\end{array}\right.
	\nn
	\ee
Comparing this relation with (\ref{jost}), we see that $\psi$ is a scalar multiple of the Jost solution $\psi_r$ with
	\begin{align}
	&R^r=R^r_{lm}:=(-1)^{l+1}\left(1+\frac{i k{f}^+_{lm}}{2\pi \zeta_{lm}}\right),
	&&T=T_{lm}:=\frac{(-1)^{l+1} i k{f}^-_{lm}}{2\pi \zeta_{lm}}.
	\label{RT=AppB}
	\end{align}
		
The total scattering cross sections measured by the observers $O^\pm$ are given by
	\be
	\sigma^\pm:=\int_{S^2}d\Omega\:|f^\pm(\vartheta,\varphi)|^2=\sum_{l=0}^\infty
	\sum_{m=-l}^l|f^\pm_{lm}|^2,\nn
	\ee
where we have used (\ref{ff=}) and the orthonormality of the spherical harmonics. According to $O^+$, 
$\sigma^+$ is the total cross section of the wave detected by her detector, while $\sigma^-$ is a measure of the power transmitted to $M^-$. Therefore, $O^+$ will identify $\sigma^-$ with the absorption cross section of the wormhole, which we denote by $\sigma$ in Section~6. 

We can use (\ref{RT=AppB}) to express $\sigma^\pm$ in terms of $R_{lm}^r$ and $T_{lm}$. This gives
	\begin{align}
	&\sigma^+=\frac{4\pi^2}{k^2}\sum_{l=0}^\infty\sum_{m=-l}^l
	|\zeta_{lm}|^2\left|R^r_{lm}+(-1)^l\right|^2,
	&&\sigma^-=\frac{4\pi^2}{k^2}\sum_{l=0}^\infty\sum_{m=-l}^l
	|\zeta_{lm}|^2\left|T_{lm}\right|^2.\nn
	\end{align}
Because for non-spherical low-energy waves $|T_{lm}|$ is negligibly small, to a very good approximation, 
	\be
	\sigma=\sigma^-=\frac{4\pi^2|\zeta_{00}|^2\left|T_{00}\right|^2}{k^2}=
	\frac{\pi \left|T \right|^2}{k^2},\nn
	\ee 
where we have employed the identity, $\zeta_{00}=Y_0^0(\vartheta_0,\varphi_0)^*=1/\sqrt{4\pi}$.

\end{document}